\documentclass[fleqn,10pt]{SelfArx} % Document font size and equations flushed left

\usepackage{lipsum} % Required to insert dummy text. To be removed otherwise

\definecolor{color1}{RGB}{0,0,90} % Color of the article title and sections
\definecolor{color2}{RGB}{0,20,20} % Color of the boxes behind the abstract and headings

\usepackage{hyperref} % Required for hyperlinks
\hypersetup{colorlinks=true,urlcolor=blue,linkcolor=blue,linkbordercolor=blue,breaklinks=true,bookmarksopen=false,pdftitle={Title},pdfauthor={Author}}
\JournalInfo{Geomagnetism and Aeronomy, 2021, Vol. 61, No. 7, pp. 1069--1074.} % Journal information
\Archive{DOI: 10.1134/S0016793221070057} % Additional notes (e.g. copyright, DOI, review/research article)
\Archive{}

\PaperTitle{
Detection of Optical Flares on the Selected G-M Dwarfs
from Long-term Photometric Series
} % Article title

\Authors{N.\,I.~Bondar'\textsuperscript{1,*}, M.\,M.~Katsova\textsuperscript{2,**}, and A.\,A.~Shlyapnikov\textsuperscript{1,***}
} % Authors

\affiliation{\textsuperscript{1}\textit{
Crimean Astrophysical Observatory, Russian Academy of Sciences, Nauchny, Russia
}} % Author affiliation
\affiliation{\textsuperscript{2}\textit{
Sternberg State Astronomical Institute, Lomonosov Moscow State University, Moscow, Russia
}} % Author affiliation
\affiliation{\textsuperscript{*}e-mail: otbn@mail.ru}
\affiliation{\textsuperscript{**}e-mail: maria@sai.msu.ru}
\affiliation{\textsuperscript{***}e-mail: aas@craocrimea.ru}
\affiliation{Received: March 6, 2021; revised: April 11, 2021; accepted: April 26, 2021}

\newcommand{\keywordname}{Keywords}
\Abstract{
We have carried out a search for flares from the analysis of light curves for 12 active G, K, and M
dwarfs. As sources of data we used ground-based observations in 2000--2020 from the photometric databases
ASAS, SuperWASP, KWS. Events of low-amplitude brightening ($\Delta V < 0.25\;$mag), which possibly could be
flares, were revealed for 11 stars. A large number of such increases in brightness were found on K dwarfs.
Events of increasing in $V$-magnitudes to 0.5~mag or more were detected on light curves of one G star, BE Cet,
and two M dwarfs. For three flares we could follow their development with time. We have estimated the duration
of these flares; they lasted more than an hour, but less than 3 hours. In most cases we could not determine
a lifetime of the suggested flares, but we believe that most of the probable flares on the investigated cool
dwarfs are short-lived, on the order of several minutes.

~

\textbf{DOI} ~ 10.1134/S0016793221070057
}

\begin{document}

\flushbottom % Makes all text pages the same height

\maketitle % Print the title and abstract box

% \tableofcontents % Print the contents section

\thispagestyle{empty} % Removes page numbering from the first page

\section{INTRODUCTION}

Starspots and flares in the atmospheres of low-mass
G-M stars are proxies of their magnetic activity
manifestation. Since the 1960s, these structures have
been studied in different spectral ranges using ground-based
and space observations. Starspots are a cause of
low-amplitude variations in brightness with the rotation
period of a star and the long-term variability,
often cyclical changes, by a few or even tens of percent
relative to the quiescent state of a star.

Although flares are associated with starspots, they
are irregular phenomena that produce significant
rapid increases of radiation in the optical continuum,
emission lines, and in X-ray fluxes. On the UV~Cet type
stars (flaring M dwarfs), most of the detected
flares are classical, i.e. with a fast rise in brightness up
to several magnitudes and a subsequent gradual decay.
The energies of such flares reach $10^{30}-10^{34}\;$erg. A
review of studies concerning flare activity on the
main-sequence G-M stars is given in the recently published
monograph by Gershberg et al. (2021). The
efforts of astronomers are currently aimed at studying
flares in the visible range of wavelengths and at searching
for powerful stellar flares.

\begin{table*}[!th] %%% Table 1
\begin{center}
%%%\vspace*{-0.9cm}
%%%\vspace*{0.2cm}
\caption{
Sample of investigated G-M dwarfs and number of positive $\Delta V$ (events of possible flares)
}
\par\medskip
\begin{tabular}{|l|l|c|c|c|}
\hline
Star & Spectral type & $V$, mag & Number of $\Delta V < 0.25^m$ & $\Delta V > 0.3^m$ \\
\hline
HD~1835, BE Cet & G3~V & 6.39 & 2 (K) & 0.3(K), 0.5 (K) \\
HD~4747, GJ~36  & G9~V  & 7.16  & 4 (A)  & -- \\
HD~97334, MN UMa  & G1~V  & 6.41  & --  & -- \\
HD~168443, NSV 24398  & G6~V  & 6.92  & 2(A), 2(K)  & -- \\
HD~17382, BC Ari  & K1~V  & 7.62  & 4 (K)  & -- \\
HD~61606, V869 Mon  & K2~V  & 7.17  & 8(A), 10(K)  & -- \\
HD~283750, V833 Tau  & K2~V  & 7.90  & 5(A), 4(S), 9(K)  & -- \\
HD~45088, OU Gem  & K3~V  & 6.75  & 1(A), 5(K) & -- \\
HD~197481, AU Mic  & M1~V  & 8.63  & 7(K)  & 0.32(K) \\
EV~Lac, GJ 873  & M4~V  & 10.26  & 1(K)  & 1.08 (K) \\
GJ~1243, KIC 9726699  & M4~V  & 12.7 $(m_{K\!p})$ & 9(S)  & -- \\
YZ~CMi, GJ 285  & M4.5~V  & 11.22  & 3(A), 7(K)  & 0.5(A), 0.67(K) \\
\hline
\end{tabular}
\end{center}
\vspace*{0cm}
{
\small
\hspace*{1.8cm}Letters A, K, S in brackets indicate catalogs ASAS, KWS, SW;\\
\hspace*{1.8cm}the magnitude $m_{K\!p}$ of the star GJ~1243 is given according to the $Kepler$ mission data.
}
\label{Table1}

\end{table*}

The white-light flares were discovered on the Sun
observing the intensity of radiation from the Sun as a
star. The contribution of optical flares can be traced
over all spectral regions; their energy distribution corresponds
to blackbody radiation with a temperature of
9000~K (Kretzschmar, 2011). Due to observations
from the $Kepler$ mission, the manifestations of activity
have been studied for a large number of solar-type and
cool stars (for instance, Davenport, 2016; Savanov and
Dmitrienko, 2017; Namekata et al., 2019). As shown
by the results of the recent statistical analysis of flares
on G dwarfs based on data of the $Kepler$ mission and
Gaia-DR2 catalog, non-steady events on solar-like
stars can reach energies within $10^{36}\;$erg (Okamoto et al.
(2021). On the modern Sun there occurred flares with
total energies not exceeding $3 \times 10^{32}\;$erg, and only the
Carrington flare on September 1, 1859 is estimated of
several times powerful. By studying optical radiation
from stars in a certain range of physical parameters, it
is possible to determine both typical and maximum
flare energies, to trace their development on G-M
dwarfs with different ages, and to investigate the distribution
of flare energies. This knowledge is necessary
for studying the formation of planetary systems and
their environment. The study of solar-like stars is
important to understanding whether superflares
were produced over the lifetime of the Sun. Here we
present the results of a search for flares on 12 G-M
dwarfs derived from the long-term ground-based photometry.
Our sample includes both poorly studied
stars and well-known ones with high-level flare activity.
The brightness of the most selected objects was
studied during the 20-year time interval, but in most
cases the data in catalogs of ground-based observational
programs have been obtained with a low-time
resolution and a short-time continuous monitoring
during a night. Therefore, our work aims to reveal
events of increase in brightness, which may indicate
the development of flares, to determine the maximum
$V$-values for possible flares and also their number for
dwarfs of different spectral types. Section 2 contains a
list of objects in our research and a description of the
used databases, Sections 3 and 4 contain processing
methods and results, a brief discussion of results is given
in Conclusions.

\begin{figure}[!th] %%% Figure 1
\centering
%%%\vspace*{-0.9cm}
\includegraphics[width=\linewidth]{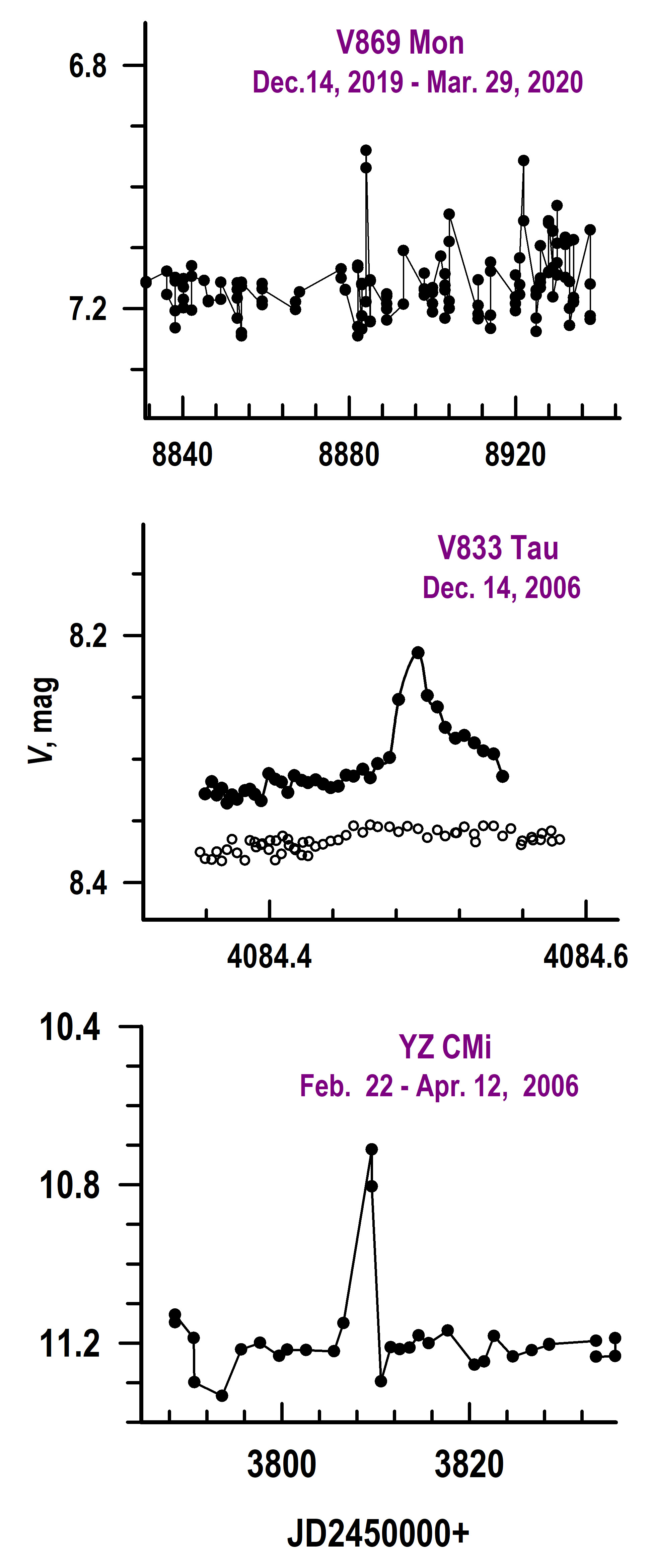}
%%%\vspace*{0.2cm}
\caption{
Examples of the selected observational series from
different catalogs and possible flares for V869 Mon
(KWS), V833 Tau (SW), open circles mark data for a comparison
star, and YZ CMi (ASAS).
}
\label{Figure1}
%\end{center}
\end{figure}

\section{OBJECTS FOR THE RESEARCH AND DATA SOURCES}

Flares on stars located on the lower part of the
main sequence are an irregular, stochastic phenomenon,
and their detection requires a systematic many hour
monitoring of stellar brightness during a night.
The long-term observational data for a large number
of stars are contained in a convenient form in photometric
databases All-Sky Automated Survey (ASAS)
(Pojmanski, 1997, 2002), the SuperWASP photometric
survey (SW) (Pollacco et al., 2006; Norton et al.,
2007; Butters et al., 2010), the Kamogata Wide-field
Survey (KWS) (Maehara, 2014).

These databases are effective sources of data for
studying starspots on the surface of G-M dwarfs, for
determining rotation periods of stars and parameters
of rotational brightness modulation, and also for
searching for cycles of photospheric activity. The light
curves constructed from these data often show short-term
increases in brightness, which are usually considered
as outliers and are excluded from data series which are analyzed
for systematic variability. Some of these brightness
increases are likely caused by real flares.

Based on the available nearly 20-year data, we can
explore whether a significant increase in brightness of
the star may be considered as flares
observed over this span. The ASAS archive can be
used to study the brightness behavior from 2000 to
2009. The best photometric accuracy was derived for
stars with magnitudes in the range of $8 < V < 12\;$mag.
The SuperWASP photometric catalog contains the
results of observations obtained in 2004--2008 for the
stars brighter than 15 mag. The accuracy of data for
stars with a magnitude of $8 < V < 11.5$ reaches 1\%, the
errors increase for brighter and fainter stars (Pollacco
et al., 2006; Norton et al., 2007).

The survey observations of Kamogata Wide-field
Survey (KWS) started in 2010 and are currently being
continued in bands $B$, $V$, $I_c$. Data series in the $V$-filter
are more numerous. The errors of observations are
given for each $V$-value, photometric precision is not
less than 5\%\ for stars with magnitudes of $5 < V < 11.5$
(Maehara, 2014).

We selected 12 G-M dwarfs, out of 4 stars in each
spectral group (Table 1), taking into account indices of
activity of stars, their stellar magnitudes, the time span
of data series in the $V$-filter in the indicated catalogs.
As we noted above, in the used catalogs errors of
observations of $V$-values increase for both faint and
bright stars, therefore we have included in our sample
stars with magnitudes in the range of $6.4 < V < 13\;$mag,
G dwarfs are the brightest among them.

Two G stars, BE~Cet and MN~UMa, are young
solar analogous (Radick et al., 2018). The sample of K
dwarfs includes active BY~Dra-type stars -- V833~Tau
and OU~Gem. The group of M dwarfs includes the
stars AU~Mic, EV~Lac, YZ~CMi, which are known by
their high-level flare activity, also the star GJ~1243, for
which a large number of flares have been found from
the extra-atmospheric observations (Hawley et al.,
2014; Silverberg et al., 2016).

\section{SEARCH FOR POSSIBLE FLARES}

For each star under study, the preliminary processing
of data series was performed. As a result, we
obtained a series of $V$-magnitudes sorted by Julian
dates, from which the random significantly fainter
$V$-values were excluded. Records in the ASAS catalog
contain for each of the five observation cameras the
observation times (Julian dates), $V$-magnitudes, and
their errors. We have compared records from different
cameras, and for further calculations there were chosen
2--3 cameras with smaller observational errors and
the best coincidence of light curves.

Then we determined average magnitudes at the
moments of observations and corresponding errors.
The full data series obtained by this way contain up to
700 $V$-values, but in the observational date 1--2 measurements
were taken usually, sometimes up to 4. Data
series in the KWS catalog include several hundreds of
measurements covering the seasonal intervals of
observations well, for a date of observation there are
given 2--4 estimates of brightness, for some dates two
exposures were made at the beginning and at the end
of the 40-min cadence.

The SW catalog contains a large number of data, up
to several thousand, obtained by several cameras with
a time resolution of 6--8 min. The time of monitoring
was several hours over a night. We examined the data
series from the SW archive and selected the sets with
the largest number of records from one of the cameras
for a more detailed analysis. Using the SW catalog we
could study only two stars, V833~Tau and GJ~1243, a
research of other chosen stars was performed based on
data of the ASAS and KWS catalogs.

We constructed light curves from data of each catalog
and found for each continuous set of observations
the mean magnitudes $V_{av}$, standard deviation $\sigma$, and
for the further flare search process we selected sets
with positive $\Delta V_i = V_i - V_{av} > 2.5\sigma$. A similar simple
statistical analysis was made in studies of data series
from the $Kepler$ mission and the ASAS-SN catalog
(Hawley et al., 2014; Martinez et al., 2020). In most
cases, increases in brightness confirmed by two or
more consecutive records of observations were considered
as possible flares. We have found clear evidence
for reality of the suspected flares for some stars, comparing
their light curves with ones for comparison
stars observed simultaneously. In addition, for the
stars with the known rotation periods, phase curves
were also constructed. This allows us to take into
account variations in the light curve produced by rotational
modulation due starspots, and also to determine
phases of an appearance of possible flares.

\begin{figure*}[!th] %%% Figure 2
\centering
%%%\vspace*{-0.9cm}
\includegraphics[width=\linewidth]{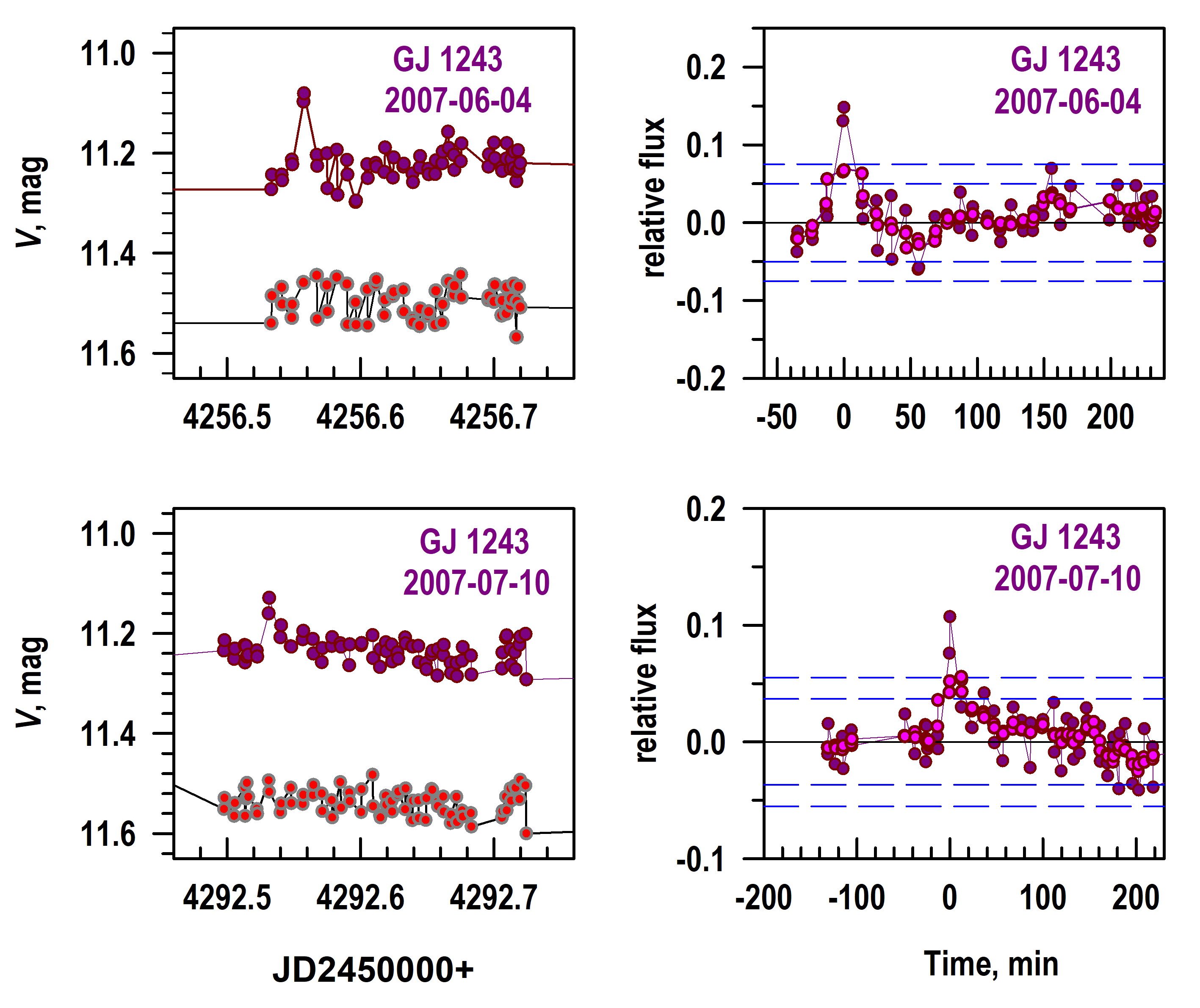}
%%%\vspace*{0.2cm}
\caption{
Two flares detected from the SW catalog data for the star GJ~1243. The left panel shows changes in brightness over time
given in Julian dates (the squares represent the comparison star), and the right panel shows the development of flares with time
in minutes from the moment of peak of brightness (black circles). The open circles show the result of five-point moving average
smoothing. Levels of $2\sigma$ and $3\sigma$ are shown by dashed lines.
}
\label{Figure2}
%\end{center}
\end{figure*}

\section{RESULTS ON IDENTIFICATION
OF FLARE EVENTS}

The events of low-amplitude ($\Delta V < 0.3\;$mag) and
short-term increases in brightness relative to the average
value in the selected interval were found for all the
stars of our sample, except for MN~UMa (G1~V). The
number of such events is given in Table 1. An increase
in brightness up to $0.3-0.5\;$mag was detected for the
stars BE~Cet (G3~V) and AU~Mic (M1~V), possible
flares of 0.5--0.67 mag were found in light curves of
the star YZ~CMi (M4.5~V). Analysis of the KWS data
for the star EV~Lac (M4~V) allowed us to define one
event of an increase in its brightness by 0.18~mag ($4\sigma$)
and one significant brightening up to 1.08~mag
detected from 4 measurements at the beginning of
observations on July 14, 2011. If this flare is real, its
duration is not less than 30~min. Figure 1 shows examples
of the light curves from different catalogs for some
stars. For stars with the known rotation periods, we
have constructed phase light curves to resolve an issue
whether the moments of appearance of possible flares
and development of starspots are consistent. Flares on
the stars BE~Cet and V833~Tau have been observed at
phases that are distant from the minimum produced by
the maximum contribution of cool starspots.

Using the SW catalog, a search for flares was performed
for two stars: for GJ~1243 (M4~V) we studied
light curves from observations on June 4, 2007 and
July 10, 2007 (Fig.~2), for V833~Tau (K2~V) we analyzed
data on December 14, 2006 (Fig.~3).

\begin{figure*}[!th] %%% Figure 3
\centering
%%%\vspace*{-0.9cm}
\includegraphics[width=\linewidth]{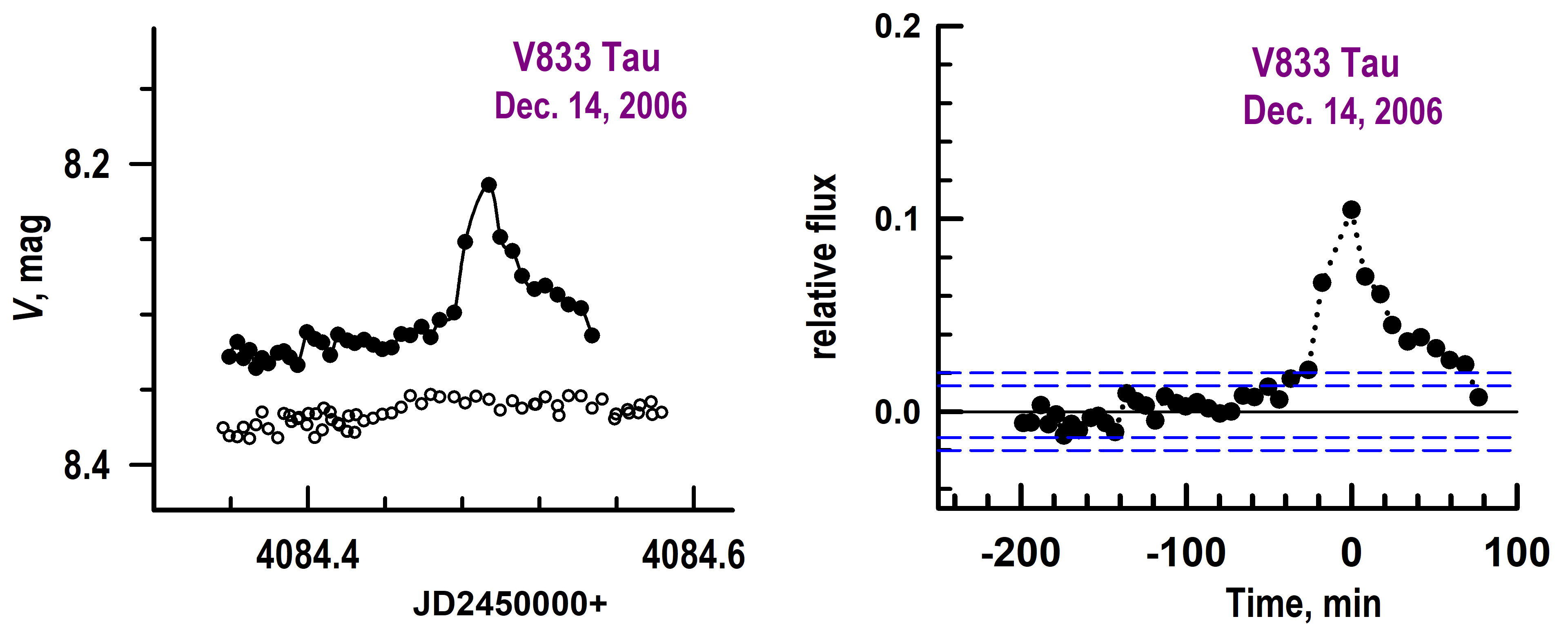}
%%%\vspace*{0.2cm}
\caption{
The same as in Fig. 2 for the star V833 Tau.
}
\label{Figure3}
%\end{center}
\end{figure*}

The presence of increases in brightness is confirmed
by the light curves of comparison stars. Figures~2 and 3
(right panels) show changes in the $V$-values over time
in minutes from the moment of maximum of increases
relative to the average value for the considered interval.
The relative intensity $f_r = f_i / f_c - 1$, where $f_i$ was 
calculated by the formula $-0.4m = log f_i$, $f_c$ is the average
value, $m = V_i$. The increase in brightness for both stars
is no more than 15\%. In order to determine the onset
and the end of a possible flare, the $f_r$-values were
smoothed by moving averages over 5 points. According
to our estimates, flares on GJ~1243 have duration
of 1.4~h and 2.4~h, and duration of the flare on V833~Tau 
was 2.2~hours.

The considered photometric catalogs contain
results of survey observations, i.e. short-term brightness
estimates, which allow one to register only those
rare flares that occur at the moment of observation.
Nevertheless, such random registrations characterize
to a certain extent the flare activity of the star. For each
catalog and each star, we determined the frequency of
brightenings (possible flares) on the value of $\Delta V <
0.25\;$mag. Its average value for each spectral type can
be considered as a characteristic of the flare activity of
stars of this spectral type.

The frequency of flares is $n = N_f / T_{obs}$, where $N_f$ is
the number of the registered flares in the observation
interval, $T_{obs}$ is the observation time of the star, $T_{obs} =
t_{exp} N$, here $t_{exp}$ is the exposure time, $N$ is the number of
$V$-data. Table 2 lists the values of $n$ for stars studied
from the KWS catalog data series. For G stars, the
average value $n$ is 0.26, for K and M dwarfs -- 0.36 and
1.06, respectively.

\section{CONCLUSIONS}

We have undertaken a search for flares on 12 active
G-M dwarfs, studying about 150 light curves constructed
from long-term photometric series of the
ASAS, SuperWASP, and KWS catalogs. Short-term
increases in brightness, which do not exceed 0.3~mag,
were detected for 11 stars, on three stars there were
revealed several possible flares with an increase in
brightness of $0.3-0.6\;$mag, one flare up to 1~mag was
suspected for EV~Lac.

The detected events of possible flares are few for
the considered 20-year spans and give poor statistics,
nonetheless the obtained results confirm the well known
conclusion that the number of flares increases
for cool stars, and their amplitude increases also. 
On K dwarfs we observed mainly low-amplitude flares.
Among several dozen events on these dwarfs, there was
no any potential flare with high amplitude.

The result on the flare activity of BE~Cet, a young
solar analog on which, in addition to low-amplitude
ones, flares up to 0.5~mag are possible, deserves further
consideration. For each star, we have studied
some series of observations, and we can note that stars
have epochs of weak and strong flare activity. Long-term
data from photometric surveys can be useful to
study changes in the level of flare activity.

Single estimates of brightness on the date of observations
stored in the ASAS and KWS catalogs do not
allow us to determine the duration of the suspected
flares, but most of them are likely short-lived and lasting
for a few minutes. Using the SW database, we traced the
development of two flares on the star GJ~1243 and one
flare on V833~Tau. According to our estimates,
their duration was within 1.4--2.4~hours.

\begin{table*}[!th] %%% Table 2
\centering
%%%\vspace*{-0.9cm}
%%%\vspace*{0.2cm}
\caption{
Estimates of the number of flares with $\Delta V < 0.25$ from the KWS data series
}
\par\medskip
\begin{tabular}{|l|l|c|c|c|c|c|}
\hline
Star & Spectral & Time interval & Number & $T_{obs}$ of star, h & Number & Number \\
& type & JD2450000+ & of data && of flares & of flare/h \\
\hline
HD 1835 & G3~V & 5548--8841 & 703 & 6.75 & 2 & 0.30 \\
HD 168443 & G6~V & 5661--9067 & 615 & 8.86 & 2 & 0.22 \\
HD 17382 & K1~V & 5527--8866 & 979 & 14.10 & 4 & 0.28 \\
HD 61606 & K2~V & 5891--8937 & 1285 & 18.50 & 10 & 0.54 \\
HD 45088 & K3~V & 5516--8550 & 1214 & 14.57 & 5 & 0.34 \\
HD 197481 & M1~V & 6457--8420 & 464 & 3.30 & 7 & 2.09 \\
EV Lac & M4~V & 5757--8451 & 638 & 9.19 & 1 & 0.11 \\
YZ CMi & M4.5~V & 6235--8573 & 608 & 5.73 & 7 & 0.87 \\
\hline
\end{tabular}

\label{Table2}
%\end{center}
\end{table*}

\section{ACKNOWLEDGMENTS}

We have used information from the SIMBAD database,
photometric data from the Hipparcos, the International
Variable Star Index (VSX) database, supported by AAVSO,
Cambridge, Massachusetts, USA, the Kamogata-Kiso-Kyoto 
Wide-field Survey. The authors are thankful to all
the staff providing the replenishment of these databases and
access to them. We would like to especially thank referees
for useful comments and suggestions.

\section{FUNDING}

These results were obtained with partial support by the
Russian Foundation for Basic Researches, project no. 
19-02-00191, and by the grant 075-15-2020-780 of the Ministry 
of Science and Higher Education of the Russian Federation.

\section{CONFLICT OF INTEREST}

The authors declare that they have no conflicts of interest.

%%%\phantomsection
\bibliographystyle{unsrt}
%\bibliography{sample}

\end{document}